\documentclass[prl,superscriptaddress,showpacs]{revtex4}
\usepackage{bm}
\usepackage{epsfig}
\begin{document}
\def\btt#1{{\tt$\backslash$#1}}
\def\tr{{\rm\,tr\,}}
\newcommand{\beq}{\begin{equation}}
\newcommand{\eeq}{\end{equation}}
\newcommand{\bdis}{\begin{displaymath}}
\newcommand{\edis}{\end{displaymath}}
\newcommand{\bea}{\begin{eqnarray}}
\newcommand{\eea}{\end{eqnarray}}
\newcommand{\barr}{\begin{array}}
\newcommand{\earr}{\end{array}}
\newcommand{\beas}{\begin{eqnarray*}}
\newcommand{\eeas}{\end{eqnarray*}}

\title{Quantum effective potential, electron
transport  and conformons in biopolymers}

\author{Rossen Dandoloff}
\affiliation{Laboratoire de Physique Th\'eorique et
Mod\'elisation, Universit\'e de Cergy-Pontoise, F-95302
Cergy-Pontoise, France}
\email{rossen.dandoloff@ptm.u-cergy.fr}
\author{Radha Balakrishnan}
\affiliation{The Institute of Mathematical Sciences, Chennai  600
113, India}
\email {radha@imsc.res.in}


\begin{abstract}

In the Kirchhoff model of a biopolymer, conformation
 dynamics can be described
in terms of solitary waves, for certain special cross-section
asymmetries. Applying this to the problem of
electron transport, we show that the quantum effective potential
arising due to the bends and twists of the polymer enables
us to formalize and quantify  the concept of a
{\it conformon} that has been  hypothesized in  biology. Its
connection to the  soliton solution of the cubic nonlinear
Schr\"{o}dinger equation emerges in a natural fashion.
\end{abstract}
\maketitle

Geometry and topology of   long chain biopolymers  such as
proteins and DNA play a significant role \cite{vino} during
processes such as replication and transcription. Interesting
experiments for studying the conformation and elastic properties
of a {\it single} polymer by bending or twisting it have been
devised \cite{smit}. The static properties of semi-flexible
biopolymers such as actin which have only bending energy, and are
described by the well-known wormlike chain (WLC) model
\cite{goldstein} with a single elastic constant, the bending
modulus. In contrast, the static DNA with its double-helix
structure is described by the wormlike rod chain (WLRC) model
\cite{fain} with an additional elastic constant, the twist
rigidity. Although considerable work has been done on various
equilibrium properties of both these elastic models, their
intrinsic dynamical properties have not been studied so far. The
latter play a crucial role in the mechanisms of energy and
information propagation along a biopolymer, an issue of vital
interest to biologists, chemists and physicists alike. The
equilibrium properties have been studied in \cite{goldstein},
\cite{fain}. The study of intrinsic static and dynamical
properties of biopolymers, taking into account their geometry, is
a subject of great importance \cite{Gor1},\cite{Gor2}. Such issues
are of vital interest in biology as well as physics, since they
would help us understand the mechanisms of storage and transport
of energy and charge along a biopolymer.

In this paper, we describe a biopolymer using the Kirchhoff model
 \cite{Gor1}. This model starts with equations that govern the dynamics
of a thin rod that in fact characterizes a polymer in the well established
WLRC model mentioned above. We are interested in its  intrinsic dynamics,
 as well as its
effect on electron transport, since the measured electrical
conductivities  of certain polymers are seen to be much larger than
expected due to conventional mechanisms.\cite{scot} Under certain conditions, polymer
conformations take on the form of spatially localized nonlinear
excitations. Applying this to the problem of electron transport,
we show that the quantum effective potential arising due to the
bends and twists of the polymer enables us to formalize and
quantify  the concept of a {\it conformon} that has been  put
forward in biology \cite{scot}- \cite{gree}. It is expected to
play an important role in statics and dynamics of biopolymers in
general. Its connection to the soliton solution of the cubic
nonlinear Schr\"{o}dinger equation emerges in a natural fashion.

We consider the biopolymer to be a very thin elastic filament (or
rod) modeled by a {\it strip}\cite{Gor2}, which is defined as a
space curve ${\bf R}(s,t)$, along with a smooth unit vector field
$\bf {d}_2$, perpendicular to the curve. Here $s$ denotes the arc
length of the polymer and $t$ is the time. The unit tangent to the
curve is given by ${\bf d}_3$  and the third unit vector of the
triad is defined as  ${\bf d}_1 = {\bf d}_2 \times {\bf d}_3$, so
that the triad $({\bf d}_3,{\bf d}_2, {\bf d}_1)$ forms a
right-handed, orthonormal frame at every point on the curve.

The space derivatives of the vectors of the frame  can be
shown to be given by the compact expression
\bea
{\bf d}_{i,s}={\bf k} \times {\bf d}_i,
\label{fe}
\eea
where $i=1,2,3$,  the subscript $s$ stands for
$\frac{d}{ds}$, and ${\bf k}$, the Darboux vector or "twist"
vector is given by
\bea
{\bf k}(s,t)= k_1{\bf d}_1 + k_2{\bf d}_2 +k_3 {\bf d}_3.
\label{darb}
\eea
Its components $k_i$, $i=1,2,3$, can be
expressed as a function of the  curvature $k$, torsion $\tau$
 and the angle $\phi$ between the
principal normal to the curve and  ${\bf d}_1$, the normal to
 the strip:
\bea
(k_1, k_2, k_3) = (k\sin\phi,~ k\cos\phi,~ \tau + \phi_s)
\label{kcomp}
\eea
Here,
\bea
k=|{\bf d}_{3,s}|
\label{k}
\eea
and
\bea
\tau={\bf d}_{3}.({\bf d}_{3,s}\times{\bf d}_{3,ss})/k^{2}.
\label{tau}
\eea

The Kirchhoff equations that govern the dynamics of the biopolymer
(modelled as a thin elastic rod)  are given (in their
dimensionless form) by \cite{Gor1}, \cite{Cole}

\bea {\bf g}_{s}= {\bf R}_{tt} \label{K1}\eea
and
\bea { \bf m}_{s}+ {\bf d}_3 \times
{\bf g} = a {\bf d}_1 \times {\bf d}_{1,tt} + {\bf d}_2 \times
 {\bf d}_{2,tt}, \label{K2}\eea
with
\bea {\bf m} = k_1{\bf d}_1 + ak_2 {\bf d}_2 + bk_3 {\bf d}_3,
\label{K3}\eea
where  the subscript $t$ stands for the time derivative $\frac{d}{dt}$.
In these equations ${\bf g}(s, t)$ and ${\bf m}(s, t)$ represent
the force (or tension) and the torque acting on each cross-section of the
 rod. The equations are obtained from the conservation of
linear and angular momentum. The parameter $a$ ($0<a\le 1$) is a
measure of the bending asymmetry of its cross-section.
 $b =2a/(1+\sigma)(1+a)$,  $\sigma$ being the Poisson ratio,
is a measure of the change in volume of the rod  as it is
stretched.

First we consider the static version of the Kirchhoff equations
(\ref{K1})-(\ref{K3}). Using the general expression \bea {\bf g} =
g_1{\bf d}_1 + g_2 {\bf d}_2 + g_3{\bf d}_3 \label {g} \eea leads
to the following system of equations \cite{FN}:

\bea g_{1,s} + k_2g_3 - k_3g_2 = 0 \label{e4}\eea
\bea g_{2,s} + k_3g_1 - k_1g_3 =
0\label{e5} \eea
\bea g_{3,s} + k_1g_2 - k_2g_1 = 0
\label{e6}
\eea
\bea g_2 = k_{1,s}
 + (b-a)k_2k_3
\label{e7}
\eea
\bea g_1 = -ak_{2,s} + (b-1)k_1k_3
\label{e8}
\eea
\bea bk_{3,s} + (a-1)k_1k_2 = 0
\label{e9}
\eea

For all $\phi =n \pi/2$, n an integer,
using eq.~ (\ref{kcomp}) in eq.~ (\ref{e9})
shows that
\bea
k_{3}=\tau=\tau_0.
\label{tau0}
\eea
Thus  the torsion of the polymer is a {\it constant}, denoted by
$\tau_0$.
As an example, we  first take $\phi=\pi$ in
eq.~(\ref{kcomp}) and analyze eqs.~(\ref{e4}-\ref{e9}).
We find $b=2a$. This implies  $a=-1/(1+\sigma)$. Further,
\bea
\mathbf {g} = a~k_{s}\mathbf {d}_1 +(a-b)~\tau_0~ k \mathbf {d}_2 +
a(-\frac{1}{2}k^2 + C_2)\mathbf {d}_3,
\eea
where $C_2$ is an integration constant. To understand its
physical significance,
note that for $k=0$ (a straight polymer), $\mathbf {g}=a~C_2\mathbf
{d}_3$. This essentially  means that  $C_2$ represents the tension
in the polymer.

With this result, eq.~(\ref{e4})
leads to the following equation for the curvature $k$:

\bea k_{ss}+\frac{k^{3}}{2}= (C_2-\tau_0^{2})k
\label{e10}
\eea

Equation (\ref{e10})) has two trivial solutions:
the straight line $k=0$,
and the circular helix $k=\sqrt{2(C_2-\tau_0^2)}$. More
interestingly, it admits  the following nontrivial solution:

\bea k=2\sqrt{C_2-\tau_0^2}~ {\rm{sech}}~\sqrt{C_2-\tau_0^2}\,s,
\label{solk}
\eea
where as already stated, $\tau_0$ and $C_2$ are constants.

For $\phi=\frac{\pi}{2}$, following the same procedure, we get
$b=2$. This implies $a=-(1+\sigma)/\sigma$. We can show that $k$
satisfies an equation of the same form as eq.~(\ref{e10}). In fact, we
can  verify that for all $\phi=n\pi/2$, $n$ any integer, the curvature
has the form given in eq.~(\ref {solk}), and as already
found, $\tau$ is just a constant, $\tau_0$.
Since the curvature must be real, eq.~(\ref{solk}) shows that
$C_2$ must be always greater than $\tau_o^{2}$.
In the case of a planar polymer,
 $\tau_0 =0$. Thus
physically, a larger tension is needed to get a twisted polymer,
for the case under discussion.

 Turning our attention to  dynamical solutions, \cite{Gor1} have noted that
 the Kirchhoff equations
(\ref{K1})  and (\ref{K2}) can support traveling wave solutions
for the curvature $k$, called {\it Kovalevskaya waves}. These are
of the same form as the static solution (\ref{solk}), where now
$s$ is replaced by $\xi=(s-vt)$,
 with $v$ the speed of these spatially localized, solitary waves,
which propagate without change of form. These arise due to a
certain nontrivial scaling property \cite {Cole} satisfied by
Kirchhoff equations.

We will consider possible quantum  mechanical implications of this
non-trivial solution for $k$, with regard to  electron transport
on a biopolymer. It has been shown by \cite{daco}, \cite{gold} and
 \cite{Clark} that a quantum particle in a thin tube whose axis
follows a space curve with curvature $k$ and constant torsion
$\tau_0$  (as in our case) {\it feels} an effective potential
\cite{Clark} of the form

\bea V_{eff}(s)=\frac{\hbar^2}{2m}[-\frac{k^2(s)}{4}+
\frac{\tau_0^{2}}{2}].
\label{veff}
 \eea

Writing down the Schr\"{o}dinger equation for an electron
in the presence of the above effective potential, and making
a gauge transformation of the wave function $\psi_1$,
by using the following appropriate  phase factor
$\psi_1(s,t)=\psi(s,t)\exp(-i\hbar\frac{\tau_0^2}{4m}t)$,
we obtain
\bea -\frac{\hbar^2}{2m}\left(\frac{\partial^2}{\partial
s^2} + \frac{k^2(s)}{4} \right)\psi(s,t)=i\hbar \frac{\partial
}{\partial t}\psi(s,t) \label{e13} \eea

After rescaling the time such that $\frac{\hbar}{4m}t \rightarrow u$ and
the coordinate $s \rightarrow \sqrt{2}s_1$,  the Schr\"{o}dinger
equation reads:
\bea
i~\psi_{u}+\psi_{s_1s_1} + \frac{k^2}{2}\psi =0,
\label{knls}
\eea
where $k=k(s_1)$, and the subscripts $s_1$ and $u$ stand
for the partial derivatives $\frac{\partial}{\partial s_1}$ and
$\frac{\partial}{\partial u}$.

Looking for solutions of eq.~(\ref{knls}) of the form
\bea
\psi(s_1,u)=k(s_1)\exp (i\alpha~u),
\label{e15}
\eea
 we get
\bea
\left( k_{s_{1}s_{1}} + \frac{k^3}{2}\right)=\alpha ~k.
\label{e16}
\eea

This equation has the same form as
eq.~(\ref{e10}), provided
\bea
\alpha = (C_2 -\tau_0^{2}).
\label{alpha}
\eea
Using the
solution given in eq.~({\ref{solk}) (with $s$ replaced by$s_{1}$)
in eq.~(\ref{e15}), we get
\bea
\psi(s_1,u)=2\sqrt{\alpha}~{\rm{sech}}~\sqrt {\alpha}~ s_1
~\exp~i(\alpha~u),
\label{brea}
\eea

where $\alpha\ge0$. It is readily seen that the wave
function of the electron
is {\it localized}  around that point on the polymer where
the maximum of its  curvature is located. Further,
it has a simple sinusoidal time-dependence
 like a "{\it breather}".

Since $\alpha =k^{2}(s=0)/4 =k_0^{2}/4$, eq.~(\ref{alpha})
leads to
\bea
\frac {k_0^2}{4} + \tau_0^{2}=C_{2}.
\label{cons}
\eea
This leads to an interesting {\it constraint} between the
maximum curvature $k_0$ and the constant
torsion $\tau_0$ of the polymer,  $C_2$ being the constant
 representing tension.

We choose various values of $\alpha$ and $\tau_{0}$ that
 satisfy this constraint.
Note that $\alpha\le C_2$, from eq.~(\ref{alpha}).
The actual conformation of the polymer
which has a space-dependent curvature
$k=2\sqrt {\alpha}~{\rm{sech}}~ \sqrt {\alpha}~ s_1$
and a constant torsion $\tau=\tau_0$,
can be found by integrating eqs.~(\ref{fe}).
Typically, we find that polymer has a single non-intersecting
twisted loop,
centered around $s_1=0$.
It straightens out as $s_1\rightarrow\pm\infty$,
as it should, since its curvature is readily seen to
vanish in those limits. Figure 1
gives an example of such a  conformation,  for $C_2=2$,
with $\alpha=1$ and $\tau_0=1$.
For the same $C_2$, smaller values of torsion, e.g., $\tau_0=0.7$,
make the loop curve more around the center,
while for  larger values, e.g.,$\tau_0=1.23$,
the opposite happens, and the loop starts "unraveling"
and straightens out more.
Our results show how the above conformation
of a polymer that emerges directly from static Kirchhoff equations,
can lead to electron localization, i.e., "trapping"  of an
electron around the
maximum curvature point on the twisted loop that develops
mid-way on the polymer.
\begin{figure}
\includegraphics{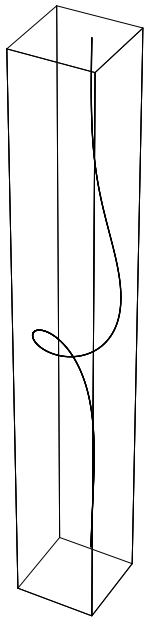} \caption{ Polymer conformation for curvature
$k$ as given in eq.~(\ref{solk}) with $(C_2-\tau_0^{2})=\alpha=1$
and torsion $\tau=\tau_0=1$. Note the localized twisted loop on
the polymer.} \label{f.1}
\end{figure}
As already  mentioned, the dynamical solutions for the curvature
$k$ are just Kovalevskaya traveling waves, given by \bea k(s_1,u)=
k(s_1-v~u)=2\sqrt {\alpha} ~{\rm{sech}} [\sqrt
{\alpha}~(s_1-v~u)]. \label{kova} \eea For this case, the wave
function of the electron is to be found as  the solution of the
corresponding time-dependent Schr\"{o}dinger equation
eq.~(\ref{knls}), where now $k=k(s_1-v~u)$ is given in
(\ref{kova}). Thus for this dynamical case, we  look for a
solution of the form \bea \psi(s_1,u)=k(s_1-v~u)~\exp~i[\lambda~
s_1 +\mu~u]. \label{e19} \eea where $\lambda$ and $\mu$ are to be
found by substituting eq.~(\ref{e19}) into eq.~(\ref{knls}), with
$k$ as in (\ref{kova}). After some algebra, we find \bea \lambda=
(v/2)~~;~~\mu=(\alpha-\lambda^{2}) =[C_2
-\tau_0^{2}-\frac{v^{2}}{4}], \label{param} \eea on using
eq.~(\ref{alpha}). Substituting for $\lambda$ and $\mu$ from
eq.~(\ref{param}) and $k(s_1-vu)$ from eq.~(\ref{kova}),
eq.~(\ref{e19}) becomes \bea \psi= 2\sqrt
{\alpha}~{\rm{sech}}{[\sqrt {\alpha}(s_1-vu)] \exp
i[\frac{v}{2}~s_{1} + (\alpha -\frac{v^2}{4})u]} \label{solit}
\eea This is identical to the {\it envelope soliton} solution of
the following, completely integrable \cite{zakh} cubic nonlinear
Schr\"{o}dinger equation (CNLSE), \bea i~\psi_{u}+\psi_{s_1s_1} +
\frac{|\psi|^{2}}{2}\psi =0. \label{cnls} \eea This is as
expected, because for the solution (\ref{e19}) that we have
considered,
 $k^{2}=|\psi|^{2}$, so that eq.~(\ref{knls}) reduces
to eq.~(\ref{cnls}).

>From eq. (\ref{solit}), it is clear that the envelope soliton
has a localized
profile: Its modulus travels with {\it envelope velocity}
$V_e=v$, while
its phase has a {\it carrier velocity} $V_c$ given by
\bea
V_c=-(\mu/\lambda)=
\frac{(v^{2}-4\alpha)}{2v}=\frac{[v^{2}-4(C_2-\tau_0^{2})]}{2v}.
\label{cv}
\eea
This leads to the well known inequality $v[v-2V_c]\ge 0$
between these two velocities
of the CNLSE soliton on the polymer. From eq. (\ref{cv}),
we see that for a given $v$,  $V_{c}$ depends on the tension
and torsion of the polymer.

Incorporating the additional phase factor $\exp (-i\tau_{0}^{2}u)$
due to the gauge transformation we had made earlier, we finally obtain
the following travelling wave solution for the wave function
$\psi_{1}(s_1,u)$:
\bea
\psi_1(s_1,u)~~=~~
2\sqrt {\alpha}~{\rm{sech}}\left(\sqrt {\alpha}~(s_1-vu)\right)\nonumber\\
\exp~i[v~s_1 + (C_2-2\tau_0^2-v^2)u],
\label{psi1}
\eea
where $\alpha$ is defined in eq.~(\ref{alpha}).
It is easy to see that in this case, the electron gets trapped
by a {\it moving}  potential well, which travels along the polymer.
To understand the conformation here, we note that the polymer now
has a  curvature which is  a Kovalevskaya solitary  wave,
traveling without change of form:
$k=2\sqrt {\alpha} {\rm{sech}} (\sqrt {\alpha}~(s_1-v~u)$.
The conformation
is again a twisted loop, but now it {\it travels}
with a constant  velocity $v$.
Thus  the electron gets trapped in the loop, and is
transported along with it,
 on the polymer. As we have seen,
its transport is soliton-like in this case.

We believe that our results provide a precise dynamical
underpinning for the {\it conformon} concept hypothesized by
various authors \cite{gree}, \cite{volk},\cite{keme} to play an
important role in biology. Green and Ji \cite{gree} state that a
conformon is a localized packet of energy (and genetic
information). It is an energy packet associated with a
conformational strain, which is localized in a region much shorter
than the length of the molecule \cite{scot}. We find the curvature
$k$ to be a localized function. Since the energy density on the
polymer is proportional to $k^{2}$, this leads to a localized
packet of energy.

Volkenstein\cite{volk} suggests that a conformon is like an
"electron plus conformational change". Kemeny and Goklany
\cite{keme} remark that "in some sense, the conformon is a
generalization of a polaron". As is well known, a polaron is a
localized electronic bound state in a discrete lattice, which is
not perfectly periodic. It  is formed by the {\it trapping} of the
electron due to the {\it nonlinearities} arising from its  strong
coupling to the lattice (phonons). Here, we mention  that in the
specific context of an $\alpha$-helical protein, starting with a
quantum mechanical {\it discrete lattice}  model, and invoking
electron-phonon
 coupling, a CNLSE has been derived in the
 continuum approximation, by Davydov\cite{davy}.

On the other hand, our work deals with conformational aspects.
Using the Kirchhoff model  (which is a continuum model {\it per
se}), we have shown that a localized electronic state arises in
the {\it curved and twisted} polymer. This is essentially because
its curvature and torsion  "interact" with the electron by
inducing a potential well, which {\it traps} it, in addition to
creating  a {\it nonlinearity} in its Schr\"{o}dinger equation.
While this scenario is indeed somewhat analogous to the  polaron
picture described above, the origin of the two mechanisms are
quite distinct, with the curved geometry of the polymer playing a
key  role in  the creation of a conformon.
 We conjecture that the moving soliton solution that arises, along with its
robust propagation can provide an explanation for the unexpectedly
high electrical conductivity (around $10^{22}$ mho/cm) found in
certain biopolymers.
 It represents  a novel mechanism of charge transport without
dissipation which is not restricted to low temperatures.

Finally, we have shown how the geometry of polymers, nonlinearity
and quantum particle transport are intimately related. Thus our
results are also likely to be of significance in other kinds of
transport phenomena in molecular biology.\\

\acknowledgments RB thanks the Council of Scientific and
Industrial Research, India, for financial support under the
Emeritus Scientist Scheme.


\begin{thebibliography}{0}
\bibitem{FN} Note that in eq.~(\ref{e8}), we have corrected
a typographical error which we found in the
corresponding equation obtained in \cite{Gor1}.
\end{thebibliography}
\end{document}